# Online Scientific Data Curation, Publication, and Archiving

Jim Gray, Microsoft Research

Alexander S. Szalay, Johns Hopkins University

Ani R. Thakar, Johns Hopkins University

Christopher Stoughton, Fermi National Accelerator Laboratory

Jan vandenBerg, Johns Hopkins University

This paper describes an approach to documenting, publishing, and preserving the Sloan Digital Sky Survey data. It is a small dataset by some standards (less than 100 terabytes), but we believe that makes it a good laboratory for thinking about the issues. Sometimes projects are so large that it is difficult to experiment and difficult to understand the whole problem.

This article first discusses generic data publication issues, and then uses the Sloan Digital Sky Survey data publication as a specific example.

## 2. What data should be preserved?

Some data are irreplaceable and must be saved; other data can be regenerated. We call the two kinds of data *ephemeral* and *stable*. Ephemeral data must be preserved, but there is an economic tradeoff between preserving it or re-computing/remeasuring stable data.

*Ephemeral* data cannot be reproduced or reconstructed a decade from now. If no one records them today, in a decade no one will know today's rainfall, sunspots, ozone density, or oil price.

The metadata about derived data products is ephemeral: the design documents, email, programs, and procedures that produce a derived dataset would all be impossible to reconstruct. But, given that metadata, the derived astronomy data can be reconstructed from the source data; it is stable. So one need only record the data reduction procedures in order to allow others to reconstruct the data.

Not all data need be saved. *Stable* data derives from simulations, from reductions of other data, or from measurements of time-invariant phenomena.

Computer simulations produce vast quantities of data. Often, one can re-run the simulation and get the answer, *if* the simulation metadata is preserved. Since computation gets a thousand times cheaper every decade, there is a tradeoff between storing the data and recomputing it. A 1990 calculation that took a year and cost a million dollars can now be done in 8 hours for a thousand dollars.

Similarly, event data for time-invariant phenomena need not be recorded. The experiment can be done again – probably more precisely and less expensively in the future based on the experiment's metadata (how it was conducted.)

In summary, *ephemeral data must be preserved; stable data need not be preserved. Metadata is ephemeral*.

**Abstract:** *Science projects are data publishers. The scale and complexity of current and future science data changes the nature of the publication process. Publication is becoming a major project component. At a minimum, a project must preserve the ephemeral data it gathers. Derived data can be reconstructed from metadata, but metadata is ephemeral. Longer term, a project should expect some archive to preserve the data. We observe that published scientific data needs to be available forever – this gives rise to the data pyramid of versions and to data inflation where the derived data volumes explode. As an example, this article describes the Sloan Digital Sky Survey (SDSS) strategies for data publication, data access, curation, and preservation.*

## 1. Introduction

Once published, scientific data should remain available forever so that other scientists can reproduce the results and do new science with the data. Data may be used long after the project that gathered it ends. Later users will not implicitly know the details of how the data was gathered and prepared. To understand the data, those later users need the *metadata*: (1) how the instruments were designed and built; (2) when, where, and how the data was gathered; and (3) a careful description of the processing steps that led to the derived data products that are typically used for scientific data analysis.

It's fine to say that scientists should record and preserve all this information, but it is far too laborious and expensive to document everything. The scientist wants to do science, not be a clerk. And besides, who cares? Most data is never looked at again anyway.

Traditionally scientists have had good excuses for not saving and documenting everything forever, it was uneconomic or infeasible. So, we have followed the style set by Tycho Brahe and Galileo – maintain careful notebooks and make them available; but, the source data is either not recorded at all, or is discarded after it is reduced.

It is now feasible, even economical to store everything from most experiments. If you can afford to store some digital information for a year, you can afford to buy a digital cemetery plot that will store it forever. It is also easy to disseminate the information either via networks or by making a copy on new media. The residual data publication costs are the costs of acquiring the data, and the costs of documenting and curating it. Storage costs are either near-zero, or soon will be near-zero. But, documenting and curating the data is certainly not free.

### 3. Who does the publication and curation?

There are several roles in the data publishing process: *Authors*, *Publishers*, *Curators* and *Consumers*.

The classical scientist (***Author***) gathers his own data, analyzes it, and submits the results based entirely on her own experiments to a journal (***Publisher***). Part of the publication task is documenting the source data so that others can use it, and documenting the processing steps so that others can reproduce them. This is onerous, but peer-review journals insist that scientists publish the data along with the results. The journals are stored and indexed in libraries (***Curators***), and read by other scientists, who can reuse the data contained in the printed journal (***Consumers***).

In the world, where data is growing at an exponential rate, much of the new data is collected by large collaborations like the Human Genome Project. These experiments take many years to build, and even longer to operate. Their data is accumulated within the project, even if it is public. Typically the data is too large to be put into a scientific journal. The only place they exist is in the project archive. By the time the data propagates to a centralized archive, newer data has arrived, swelling the overall data volume. Thus, most of the data will be still owned by the projects.

Unwillingly, and sometimes unknowingly, projects become not only Authors, but also Publishers and Curators. The Consumers interact with the projects directly. Scientists are familiar with how to be an Author, but they are just starting to learn, out of necessity, how to become a Publisher and Curator. This involves building large on-line databases and designing user interfaces. These new roles are turning out to be demanding and require new skills.

Instruments like the Large Hadron Collider at CERN and the Sloan Digital Sky Survey produce data used by a large community. Building and operating the instrument and its processing pipeline is a specialty – other scientists use the data that the instrument-builders gather. Many scientists combine data from different sources and cross-compare them. One sees this in astronomy, but the same phenomenon occur in genomics, in ecology, and in economics.

So, there is social pressure on data gatherers to publish their data in comprehensible ways and there is demand for these data publications. But the actual data publication process is onerous. The two central problems are:
**Few Standards**: There are few guidelines for publishing data. There are fewer metadata standards. What standards there are, are not widely used. The publisher must select or invent his own: deciding on units, coordinate systems, measurements, and terminology. Metadata is often done as best-effort design documents.
**Laborious**: It is laborious to document the data and the data reduction process. There are few tools. The reward system does not recognize its value. Rather the documentation is a pre-requisite to publishing the science results.

As bleak as this picture sounds, most scientific groups have carefully documented and published their data. The Genomics community [NCBI] is one example, and the Astronomy community [FIRST, ROSAT, DPOSS] give others. These groups have had to invent their own standards: deciding on units, coordinate systems, terminology, and so on. They have had to do the best they could in documenting metadata.

The astronomy community has launched the Virtual Observatory effort as an attempt to overcome both the standards problem and to make it easier to publish scientific data. Establishing a critical mass of publishers all using a common set of standards, will make it easier for the next publisher to decide what to do. Building tools that make it easy to document metadata will pioneer a new form of publishing, much as Tyco Brahe and Galileo did.

Astronomers will likely reinvent many of the concepts already well developed in the library and museum communities. Librarians would describe documenting the metadata as ***curating*** the data. They have thought deeply about these issues and we would do well to learn from their experience. Curation is an important role for Astronomy projects, and it is central to the design of the Virtual Observatory.

### 4. Who does the preservation?

When first published, data is best provided by the source, though in some disciplines it is also registered in a common repository (e.g. in Genomics, GenBank registers new sequences.) But in Astronomy, the derived data products evolve over time as the science team better understands the instrument. So, it is generally best to go to the data source while the project is underway.

Longer term (years) the data should be placed in an archive that will preserve and serve the data to future generations. This archive function is different from traditional science project functions and so is better done by an organization designed for the task. Ideally, the data is recorded in several archives in several locations so that the data is protected from technical, environmental, and economic failures. These archives will form the core of the Virtual Observatory, but it is likely that there will be disproportionate interest in the "new" data that has not yet moved to the archive.

### 5. Sloan Digital Sky Survey as a case study

The Sloan Digital Sky Survey is using a ground-based telescope to observe the ¼ of the celestial sphere over a 5 year period. It will observe about 300 million objects in 5 optical bands and measure the spectra of a uniform million-galaxy sub-sample. Observational data is processed through a sophisticated software pipeline and is available for scientific study about 2 weeks after it is acquired.

## 5.1. Units, measurements, formats

The first question to ask of any data publication project is "What shall we publish?" Beyond the raw pixel data coming from the telescope, what data products should the project produce? This is largely a science question, but once the metrics are chosen, the next question relates to how the metrics will be named, what units will be used, how errors will be reported (e.g. there should be a standard-error estimate with each value) and what are the data formats?

There are some standards in this area, the metric system, the World Coordinate System, and some IAU standards for names based on sky positions, and some units. For example, fluxes are often measured in logarithmic units (magnitudes), but radio astronomers prefer linear fluxes measured in Jansky. They have the same meaning, they are both standards, but still conversions are required. There are also many established and some emerging data representation standards that sound like alphabet soup, FITS, XML, SOAP, WSDL, VOtable, … Each project must pick its own units and definitions and shop among these standards. This is an active area of discussion in the Virtual Observatory Forum [VOforum]. The hope is that consensus will emerge in the years to come.

## 5.2. Editions and the data pyramid

The first SDSS public data installment, about 5%, called the Early Data Release (EDR) was published in June 2001. The next installment will appear in early 2003 and will comprise about 30% of the survey. The data is published on the Internet, along with its metadata and documentation.

We call a particular publication, an *edition*. Each edition adds new data and corrects problems discovered in the previous edition – typically bugs in the pipeline programs or procedures. All the edition's data is processed in a uniform way: the old data is reprocessed with the new software and the new data is processed with the same software.

One might think that the newer edition completely replaces the previous edition – but that is not so. Once published an edition should be available forever. There are two reasons for this: the short term need of scientists to continue their work on the old dataset, and the long-term need for the data to be available so that scientists can reproduce and extend any published work based on that data.

The day a new edition appears, some scientists will be in the midst of studies using the "old" edition. Shifting to the new edition might introduce inconsistencies in their analysis; and at the least it will require some re-testing of previous work. So, data publication must be structured to allow scientists to convert at their convenience, not the convenience of the publisher.

Since scientific work is based on a particular edition, it is important that that edition be available so that subsequent experiments can be done to confirm the published science results and to experiment with alternatives.

Consequently, each edition must remain available forever. This gives rise to a *data pyramid*. If there are $N$ editions, then there will be $N$ copies of the first edition's data, $N-1$ copies of the second, and so on. The sum of this series is approximately $N^2/2$.

This quadratic data growth is not a serious problem for the SDSS. Most of the bytes are pixel source data (level-1A using the NASA EOS terminology [EOS]). Only the derived data products (level 2) change from one edition to the next. The derived products are five times smaller than the level-1A data. Hence the four SDSS editions stored as a data pyramid are likely to double the survey's storage needs. If storage prices continue to drop as they have been, this will not be a problem.

If the data pyramid becomes too large, one strategy is to either not store old version of derived data (they can be reconstructed from the metadata and level-1A data on demand), or to compress the data (storing only the differences) as suggested by [Buneman].

## 5.3. Data inflation

The data pyramid increases storage demands, but also there seems to be a tendency for derived data products to proliferate – dubbed *data inflation.* For SDSS derived data products of an edition are about five times larger than the core catalog size. The data pyramid doubles this so that in the end, the derived data products will be about ten times larger than you might guess.

The second edition of the SDSS survey has about 5 terabytes of observational data (pixel data) that can be lossless compressed in half so that it will occupy 2.5TB. The edition also has about 400 gigabytes of catalog data derived from the pixel data. This catalog has the attributes of 100 million photographic objects and 300,000 spectra.

This suggests that the next edition will need about 2.5TB of compressed pixel storage for the source data, and .4 TB for

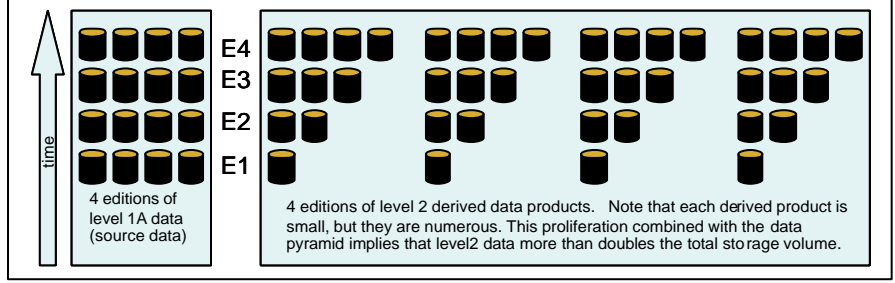

Figure 1: Data inflation is caused by the data pyramid and by a proliferation of derived data products. The data volume doubles for the SDSS.

4 editions of level 1A data (source data)

4 editions of level 2 derived data products. Note that each derived product is small, but they are numerous. This proliferation combined with the data pyramid implies that level2 data more than doubles the total storage volume.

the catalog data. But data inflation expands the catalog data to nearly 2.4 TB. First, the catalog data nearly doubles in size when it is placed in an SQL database and heavily indexed. Second, the science mission requires that the data be stored in three-ways: (1) the way it was gathered (a *runs* database), (2) the way it was used for spectrographic target selection (*target* database), and (3) the final data products (called the *best* database). So, the 400GB of derived data has inflated to about eight fold to 2.3 TB. When we add the data pyramid the catalog swells to 2.4 TB. Consequently the next edition of the SDSS data will be approximately 5 TB.

Ultimately we expect the compressed level-1A (pixel-level) SDSS data to be about 10TB, and the final data pyramid to be another 10TB, of which the final best-catalog will be 2TB. To summarize, the SDSS derived data is about ten times smaller than the compressed level-1A data, but the data inflation creates derived data (level-2) comparable to the size of the level-1A data.

### 5.4. Curation: Capturing the ephemera

Data is incomprehensible and hence useless unless there is a detailed and clear description of how and when it was gathered, and how the derived data was produced. It is difficult and tedious to document all the design decisions that go into a system. Interestingly, now most systems are designed and built by groups that are widely distributed, and who communicated primarily electronically (articles, email, and teleconferences). It is fairly easy to capture and archive all this information.

All versions of programs are archived in a source-code control system. Design documents are all electronic and are posted at various serves. Suggestions for changes, questions, and bug reports are posted at a bulletin board and tracked by a database. These discussions contain a wealth of information. The observatory generates detailed event logs as the data is gathered. The pipeline processing and publication processes also generate detailed logs describing how products were produced. The peer-review process and data-use process generates both commentaries and questions and answers. The answers contain a wealth of detailed design information.

Each project scientist has a small archive of notes, memoranda, and experiments. If consolidated, these unpublished archives could be indexed by a full-text search system and might be useful in the future.

The SDSS consortium recognizes these curation issues, and is moving to curate these ephemera and metadata; but, it is in the early stages of that process. The project still has 4 years to go, but it recognizes that the data publication must contain as much of this metadata as possible.

As scientists use the data, sometimes they find small errors, related to one object or database row. They might make an "annotation" in their own notebook about it, like "do not use this, it is just a star superposed on a galaxy". It will be extremely useful to systematically capture these annotations; so that others, if in doubt, can benefit form them.

### 5.5. Data publication and access

Once an edition is ready, how can scientists find and access the data? Should we get an ISBN number from the Library of Congress and register the edition at Amazon.Com?

Today projects are responsible for publishing their data. The project sets up a web site and offers the data to any and all. Users find the data through word-of-mouth or from references in the literature, or by searching portals [NED, VizierR, Simbad, SkyView]. The Virtual Observatory is likely to unify this process and make it easy for scientists to register new datasets and for users to find these datasets.

Traditionally, data has been published as files organized in directories that are named to give temporal and spatial clues about the file's contents. The files are in a self-defining format (e.g., [FITS]) with additional metadata.

To get a subset of the data, a scientist copies all the relevant files via the internet or requests them via public post. Then the scientist reads the files and subsets them as desired. This mail-order or file-transfer model fails when datasets reach terabytes: a scientist wanting a color-cut of the SDSS database would have to download several terabytes and then discard the 99% of the data not relevant to the query.

Projects now publish their data as databases that allow users to request personalized subsets of the data. The requests can be sent over the internet and the replies returned to the scientist. Again, the Virtual Observatory will carry this idea much further, allowing users to request data from many archives at once, and returning relevant data from each of them. This is evolving rapidly, for example see http://skyserver.sdss.org/en/tools/search, or the more recent http://skyquery.net/

In addition to the published editions, ***pre-publications*** are available to the domain experts who validate the data as it is loaded and scrubbed. Access to this pre-publication edition is limited to the peer-review committee.

When a user wants a large fraction of a database, more than 10% or more than 10GB in today's technology, it is more convenient and more economical to send the user a clone of the database -- terra-scale sneaker-net [Gray]. The clone is an inexpensive computer with attached disks, a resident database system, and a high-speed network interface. Currently, an SDSS database clone costs about two thousand dollars. Even though the database size is growing quickly, storage prices are declining even more quickly. The full SDSS clones will likely cost less than ten thousand dollars in 2006. This makes it economical to have copies in many places, both for easy access and for archival protection.

### 5.6. Preservation formats and format conversion

Science projects eventually come to an end and the researchers move on to new instruments and experiments. But, the data needs to be available to future researchers. Even today, some scientists are looking at the astronomy records of 50 and 500 years ago – mostly to do temporal comparisons.

Science datasets must eventually transition to archive facilities that can preserve the data. The archive needs to store the data, provide access to it, and to translate the data and software forward in time so that new applications can access the old data. With improved tools and hardware, a dataset's maintenance cost drops about 100x per decade. So, the eternal archive cost for the data should be less than 5% of the first-year data service cost.

A decade ago, 100 GB was considered a huge database. Today it is about ½ of a disk drive and is quite manageable. We assume that 10-year old databases will always seem easy–to-manage. For example, a thousand 1990's magnetic tapes fit on a single online disk today – so it is both economical and desirable to bring the old data forward and store it on newer technology. Modern tools automatically "ingest" flat files and place them in databases. Moving from one database to the next is not a huge problem. The SDSS converted from one to another with modest effort.

To be safe, the data must be archived by two or more independent organizations in different countries, so that it can survive natural and political disasters. This also gives better access for users in Europe, Africa, Asia, Australia, and South America.

### 6. Summary

Much of what is described here is implicit or explicit in current projects: To summarize:

Ephemeral data must be preserved; stable data need not be preserved. Metadata is ephemeral and allows the stable data to be reconstructed.

Data publication is really data curation. The curators must capture as much metadata as possible. Project design documents, discussions, procedures, software, and operations logs are part of the metadata and should be part of the data publication. The science community can benefit from digital library research in this area.

Building consensus around a few central standards for data representation and publication is one of the main contributions the Virtual Observatory can make. As it stands, each project has to remake these decisions today.

Data is published in editions that must remain available forever. These editions create a data pyramid and data inflation.

Users cannot reasonably look at all of the data on their workstations, so the system must include a query interface that lets users extract personalized subsets of the data that can then be downloaded over the Internet. The VO promises to unify all these databases and give personalized cross-correlated subsets derived from many archives.

Ultimately, projects end, and the data need to reside in a long-term storage facility, a science archive. Among other things, this facility must keep the data accessible by transforming it to new media and new data formats.

The Virtual Observatory effort is working to address all these issues—trying to build consensus around standards, trying to make it easy to publish data, trying to make it easy to find data, and trying to build a critical mass of cooperating astronomy archives.

### 7. Acknowledgements

We would like to acknowledge the support of Compaq/HP for providing the hardware for the SkyServer. AS would like to acknowledge support from NASA and the NSF.